\documentclass[12pt]{article}

\def\pbnr{FERMILAB-CONF-13-CC/T}
\def\speaker{Andreas Kronfeld}
\def\onbehalfof{}
\def\title{Theoretical Perspective of Charm Physics}
\def\affiliation{Theoretical Physics Department \\
Fermi National Accelerator Laboratory,\SupportedBy{\support} Batavia, Illinois, USA}
\def\support{Operated by Fermi Research Alliance, LLC, under Contract No.~DE-AC02-07CH11359 with 
the US Department of Energy}

\newcommand\pubnumber{\pbnr}
\newcommand\pubdate{\today}
\def\Title#1{\begin{center} {\Large #1 } \end{center}}
\def\Author#1{\begin{center}{ \sc #1} \end{center}}

\newcommand{\OnBehalf}[1]{\sbox0{#1}\ifdim\wd0=0pt
        {}
	\else
	{\\on behalf of #1}
	\fi}
\newcommand{\SupportedBy}[1]{\sbox0{#1}\ifdim\wd0=0pt
        {}
	\else
	{\footnote{#1}}
	\fi}
\def\Address#1{\begin{center}{ \it #1} \end{center}}

\newcommand\pubblock{\includegraphics[width=5cm]{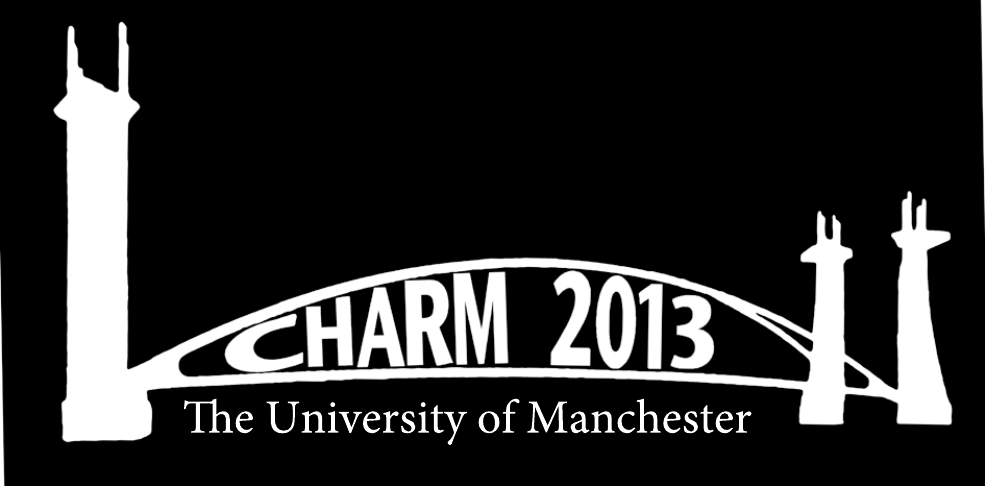}\hfill{\begin{tabular}{l} \pubnumber\\
         \pubdate  \end{tabular}}}
\newenvironment{Abstract}{\begin{quotation}  }{\end{quotation}}
\newenvironment{Presented}{\begin{quotation} \begin{center} 
             PRESENTED AT\end{center}\bigskip 
      \begin{center}\begin{large}}{\end{large}\end{center} \end{quotation}}
\def\Acknowledgements{\bigskip  \bigskip \begin{center} \begin{large}
             \bf ACKNOWLEDGEMENTS \end{large}\end{center}}
\def\venue{The 6$^{th}$ International Workshop on Charm Physics\\
(CHARM 2013)\\
Manchester, UK,  31 August -- 4 September, 2013}

\def\beq{\begin{equation}}
\def\eeq#1{\label{#1}\end{equation}}
\def\eeqn{\end{equation}}

\def\beqa{\begin{eqnarray}}
\def\eeqa#1{\label{#1}\end{eqnarray}}
\def\eeqan{\end{eqnarray}}

\let\littlebar=\bar
\let\bar=\overbar

\def\Dslash{\not{\hbox{\kern-4pt $D$}}}
\def\dslash{\not{\hbox{\kern-2pt $\del$}}}

\def\msb{{\bar{\ssstyle M \kern -1pt S}}}

\newcommand{\text}[1]{\mathrm{#1}}

\usepackage{bm}
\usepackage{graphicx}
\usepackage{hyperref}

\begin{document}
\begin{titlepage}
\pubblock

\vfill
\Title{\title}
\vfill
\Author{\speaker\OnBehalf{\onbehalfof}}
\Address{\affiliation}
\vfill
\begin{Abstract}
A perspective on charm physics, emphasizing recent developments, future prospects, and the interplay with
lattice~QCD.
\end{Abstract}
\vfill
\begin{Presented}
\venue
\end{Presented}
\vfill
\end{titlepage}
\def\thefootnote{\fnsymbol{footnote}}
\setcounter{footnote}{0}

\section{Interplay of Charm and Lattice QCD}

A talk opening a conference should give participants something to chat about while slurping coffee, touring a
stately home, or dining at the Old Trafford.
This talk is supposed to look at the theory of charm physics.
I'd like to do so through the lens of a narrative, which was developed as port of the CLEO-$c$
proposal~\cite{Briere:2001rn}, that experimental charm physics can and should be used to validate lattice QCD.
This story is now twelve years old, and it is time to ask whether it is still as useful (scientifically) in
2013 as in~2001.

Ths first thing to keep in mind is that the least familiar aspects of numerical lattice gauge theory are very
well tested.
Monte Carlo integration codes are checked numerous ways, such as reproducing expansions around the strong-
and weak-coupling limits, writing two separate programs, and, nowadays, carrying out unit tests of individual
modules.
The codes that compute quark propagators have always had a simple built-in test.
They solve a big matrix problem of the form $MG=b$, and the solution~$G$ can always be muliplied with the
Dirac matrix $M$ to check whether~$b$ is reproduced.
In fact, lattice QCD practitioners were in a position to pioneer scientific usage of
GPUs~\cite{Egri:2006zm,Babich:2011np}, because this check makes it possible to use gamers' cards that don't
implement error-correcting arithmetic.

There are also several theoretical properties of gauge theories than can be tested without reference to
experiment.
For example, the eigenvalues of the Dirac matrix satisfy certain theorems~\cite{Leutwyler:1992yt}.
The hep-lat arXiv contains numerous papers exploring this connection, far too many to cite here; for a review
of one slice of this work, see Ref.~\cite{Splittorff:2012hz}.
A bit closer to experiment, observables such as the pion mass must depend on the quark mass in a way
consistent with spontaneously broken chiral symmetry.
Monitoring his behavior is a routine part of lattice QCD.

Finally, lattice gauge theory starts with a mathematically sound footing to define continuum gauge theory.
To interpret numerical data at nonzero lattice spacing and at finite volume, we use the same kind of 
effective field theories used throughout theoretical physics.
Once the numerical data have been generated, any well-trained theorist should be able to fit them to an EFT 
formula and learn something from the~fit.

Most of these points could have been made twelve years ago, so what has changed? First, most lattice-QCD
calculations of the twentieth century were marred by something called the quenched approximation, in which
the dynamical effects of quark loops are omitted, and approximated very roughly with shifts in the bare
parameters.
The error associated with this approximation is difficult to estimate.
In addition, the up and down quark masses in the computer took values around that of the strange quark, which
stymied the chiral tests mentioned above.
In 2001, algorithms and computers began to overcome these obstacles.
The time was ripe to try to compute with lattice QCD some observables related to charm physics: the
calculations were feasible; the correct results were not known experimentally; and experiments were about to
improve the measurements, especially CLEO-$c$, BaBar, Belle, and CDF.
These calculations could test the foundations, the methodology, and---not least---the practitioners.

After several successful postdictions in 2003~\cite{Davies:2003ik}, a collaboration of collaborations (of
which I am a member) used the same methods (the fastest ones) to predict semileptonic form factor for 
$D\to Kl\nu$~\cite{Aubin:2004ej}, charmed-meson decay constants~\cite{Aubin:2005ar}, and the mass of $B_c$
meson~\cite{Allison:2004be}.
These were all quickly confirmed by experiment.
Meanwhile, all of these calculations have been updated and extended: the normalization~\cite{Na:2010uf} and
shape~\cite{Koponen:2013tua} of the form factors, and the $B_c$ mass with an untested prediction of the
$B_c^*$ mass, $M_{B_c^*}=6.330(9)~\text{GeV}$~\cite{Gregory:2009hq,Gregory:2010gm}.
(References to updates of decay constants are given in Sec.~\ref{sec:leptonic}.)  %
Meanwhile, there are numerous other results of general interest (reviewed in Ref.~\cite{Kronfeld:2012uk}),
including calculations of the baryon mass spectrum to 2--4\%~\cite{Kronfeld:2012uk,Fodor:2012gf}.
Indeed, as discussed in other talks at this conference, lattice-QCD spectroscopy has advanced to
excited~\cite{Padmanath:2013bla} and exotic~\cite{Prelovsek:2013cta} states.

In light of this progress, we should ask ourselves several questions.
Where do inexorably smaller lattice-QCD errors lead? %
Does it make sense to mount new experiments just to test the results of the (much less expensive) computers?
What is the scope---and where are the challenges---of lattice QCD? %
Many interesting problems are precisely those that cannot be validated, because theory and experiment are in
conversation.
Tests of QCD, just like tests of QED, have their place, but it is more interesting to ponder the muon's
anomalous magnetic moment as a probe of new physics rather than a test of QED.
It is somewhat amusing to hear (from some experimenters) that simple matrix elements (``gold-plated'' in the
sense of Ref.~\cite{Davies:2003ik}) still need validation, while very very messy things, like the structure
of the $XYZ$ states, will be understood only with lattice QCD~\cite{Braaten:2013oba}.

The following sections survey recent developments in charm physics with this narrative in mind.
I think that its utility has run its course, and that there are more interesting facets to discuss.
I'll talk about the charm mass and its role in Higgs physics (Sec.~\ref{sec:higgs}),
leptonic and semileptonic decays (Sec.~\ref{sec:leptonic}), and $D^0$-$\littlebar{D}^0$ mixing and
nonleptonic decays (Sec.~\ref{sec:nonleptonic}).
Section~\ref{sec:outlook} provides an outlook.

\section{Charm and the Higgs Boson}
\label{sec:higgs}

The biggest news in particle physics since Charm 2012 is the observation at the LHC of a new particle with
mass 126~GeV \cite{Aad:2012tfa,Chatrchyan:2012ufa}.
As measurements of its properties improve, it is beginning to look a lot like a standard Higgs
boson~\cite{Dobrescu:2012td}.
The measured mass lies in the region for which the standard Higgs boson has many measurable
branching ratios, including $H\to c\littlebar{c}$.
This decay is interesting, because a comparison of the measured branching ratio with the standard-model
expectation tests the hypothesis that the standard Yukawa couplings generate the mass of up-like quarks.
In the standard model, $\textrm{BR}(H\to c\littlebar{c})\approx 2\times10^{-2}$, with
$M_H=126$~GeV~\cite{Denner:2011mq}.

The essential point of contact with charm physics is the charmed quark mass: the very nature of fermion mass
generation via Higgs bosons is that the branching ratio is proportional to $m_c^2$.
For the standard Higgs boson, the branching ratio is also approximately proportional to $m_b^{-2}$, because
$H\to b\littlebar{b}$ is the dominant branch.
The most precise determinations of both masses comes from measuring the quarkonium correlator, in the
timelike region, in $e^+e^-$ collisions near $Q\littlebar{Q}$ threshold, or from computing the same
correlator, in the spacelike region, with lattice QCD.
The latest results are (in the $\overline{\rm MS}$ scheme) $m_c(m_c)=1.279(13)$~GeV \cite{Chetyrkin:2009fv}
and $m_c(m_c)=1.273(6)$~GeV \cite{McNeile:2010ji}, respectively.
For more details, see Ref.~\cite{Davies:2013dga}.
Note that the Particle Data Group~\cite{Beringer:1900zz} acknowledges the lattice-QCD result as the most
accurate, as it does for~$\alpha_s$.

\section{Leptonic and Semileptonic Decays}
\label{sec:leptonic}

\begin{figure}[b]
    \centering
    \includegraphics[width=0.48\textwidth]{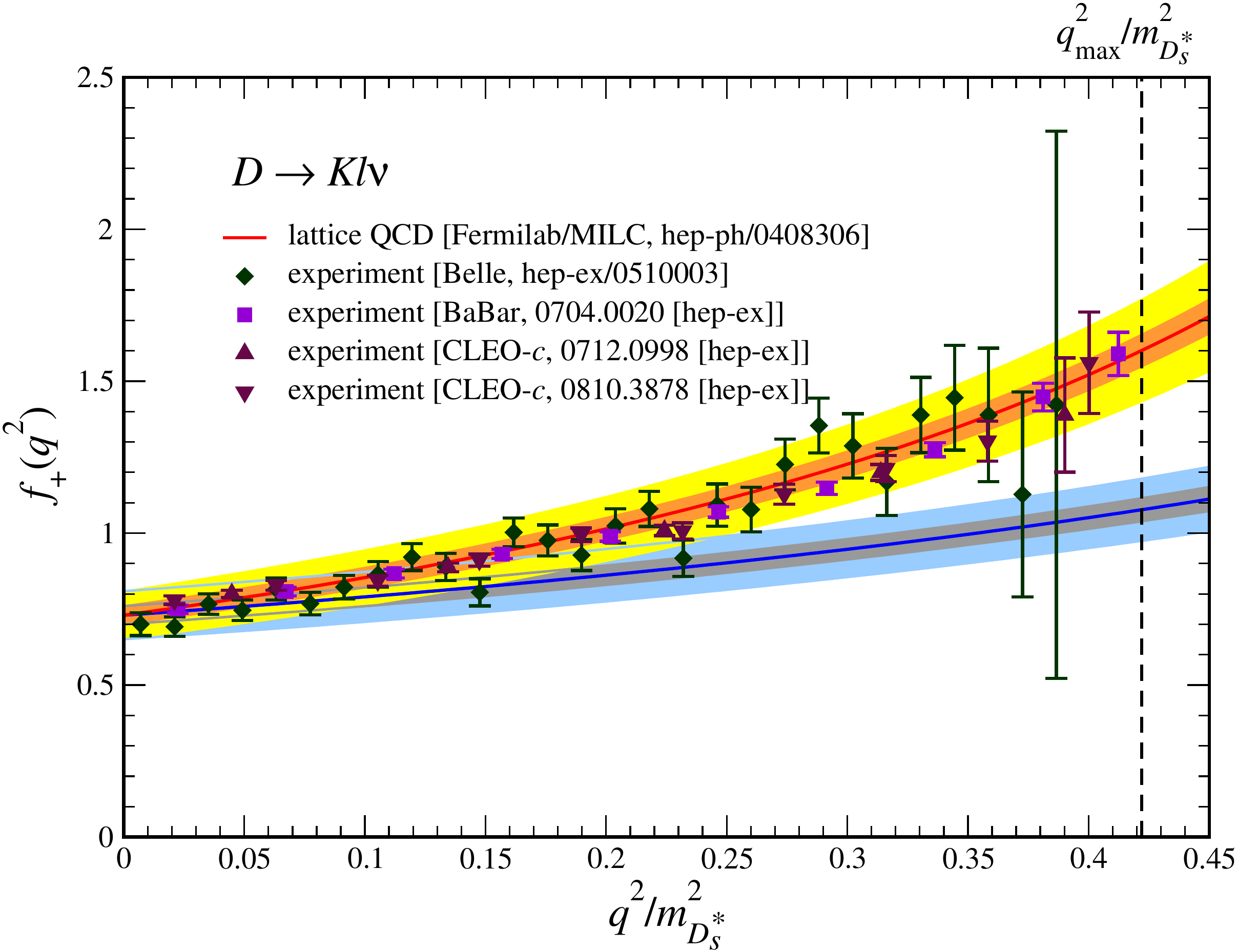}\hfill
    \includegraphics[width=0.48\textwidth]{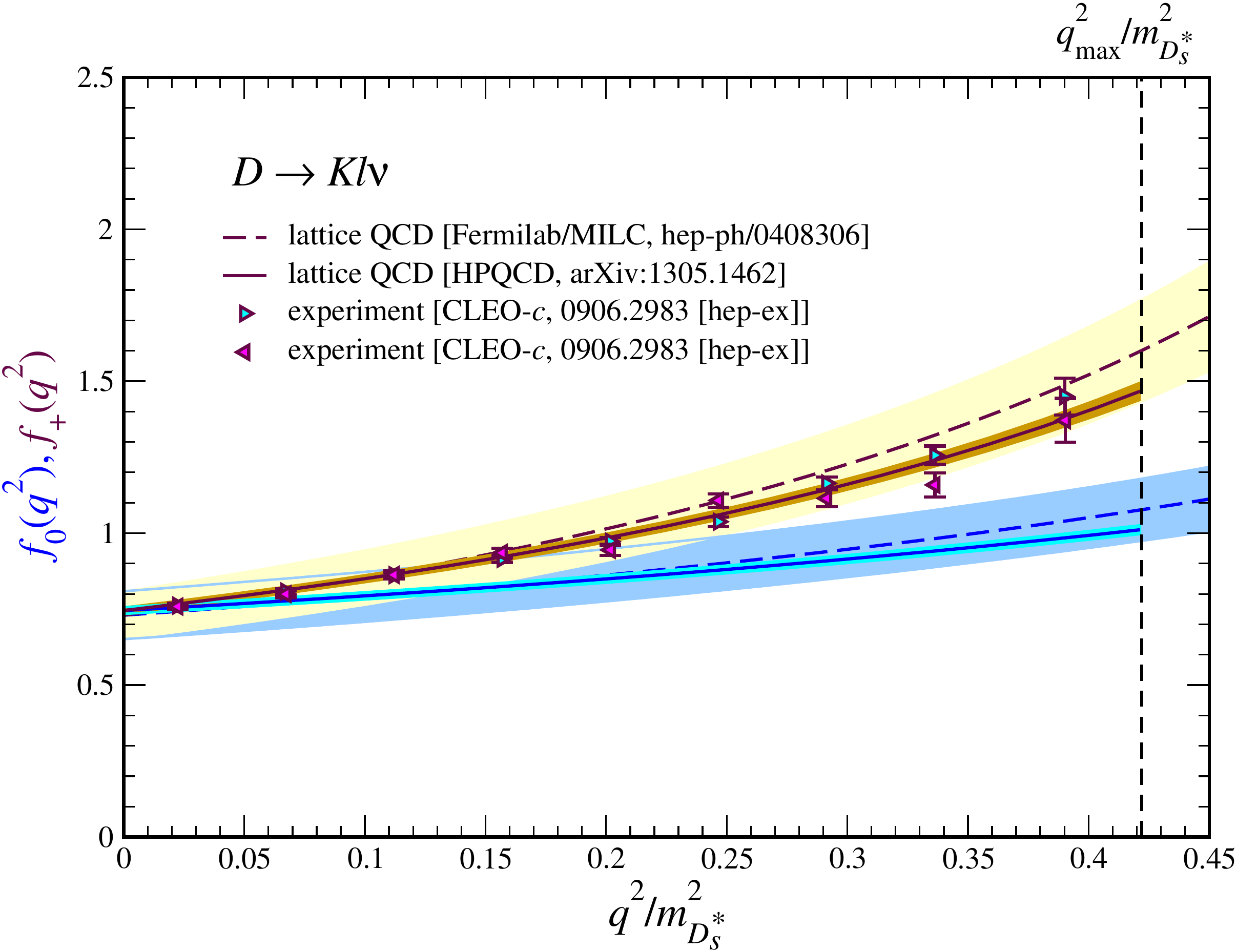}
    \caption{Semileptonic form factors, then and now.}
    \label{fig:ffDK}
\end{figure}

Semileptonic decays offer an example of the past decade's developments.
The plot on the left of Fig.~\ref{fig:ffDK} shows the 2004 calculation of the $D\to Kl\nu$ semileptonic form
factors~\cite{Aubin:2004ej,Bernard:2009ke}, together with measurements from Belle~\cite{Abe:2005sh},
Babar~\cite{Aubert:2007wg}, and CLEO-$c$~\cite{CroninHennessy:2007aa,Ge:2008aa}.
To make the comparison, the measured $|V_{cs}|f_+(q^2)$ is divided by $|V_{cs}|$ as deduced from CKM
unitarity.
The plot on the left was widely hailed as a validation, and provided confidence that similar form factors
could be used to determine $|V_{ub}|$ from $B\to\pi l\nu$ decay.

CLEO-$c$'s final results~\cite{Besson:2009uv} land on the lower edge of the error band from lattice QCD 
\emph{anno}~2004.
This development is shown on the right of Fig.~\ref{fig:ffDK}: the data lie near the bottom of the 
straw-colored band.
A~2013 calculation~\cite{Koponen:2013tua} (brown band), which is both more precise (higher statistics) and more accurate 
(smaller systematics), goes straight through the experimental points.
The systematic improvement stems from small-enough lattice spacing so that charm quarks are
now ``light'' quarks, not ``heavy'' quarks.
One now follows the strategy of $B$ decays, determining $|V_{cs}|$ from a combined fit
to lattice-QCD and experiment~\cite{Koponen:2013tua}.
The agreement of the experimental data with the new lattice-QCD curve tells us that the unitarity tests of
the second row and second column are well satisfied.
For further discussion, see Ref.~\cite{Koponen:2013ila}.

\begin{figure}[b]
    \centering
    \includegraphics[width=0.6\textwidth]{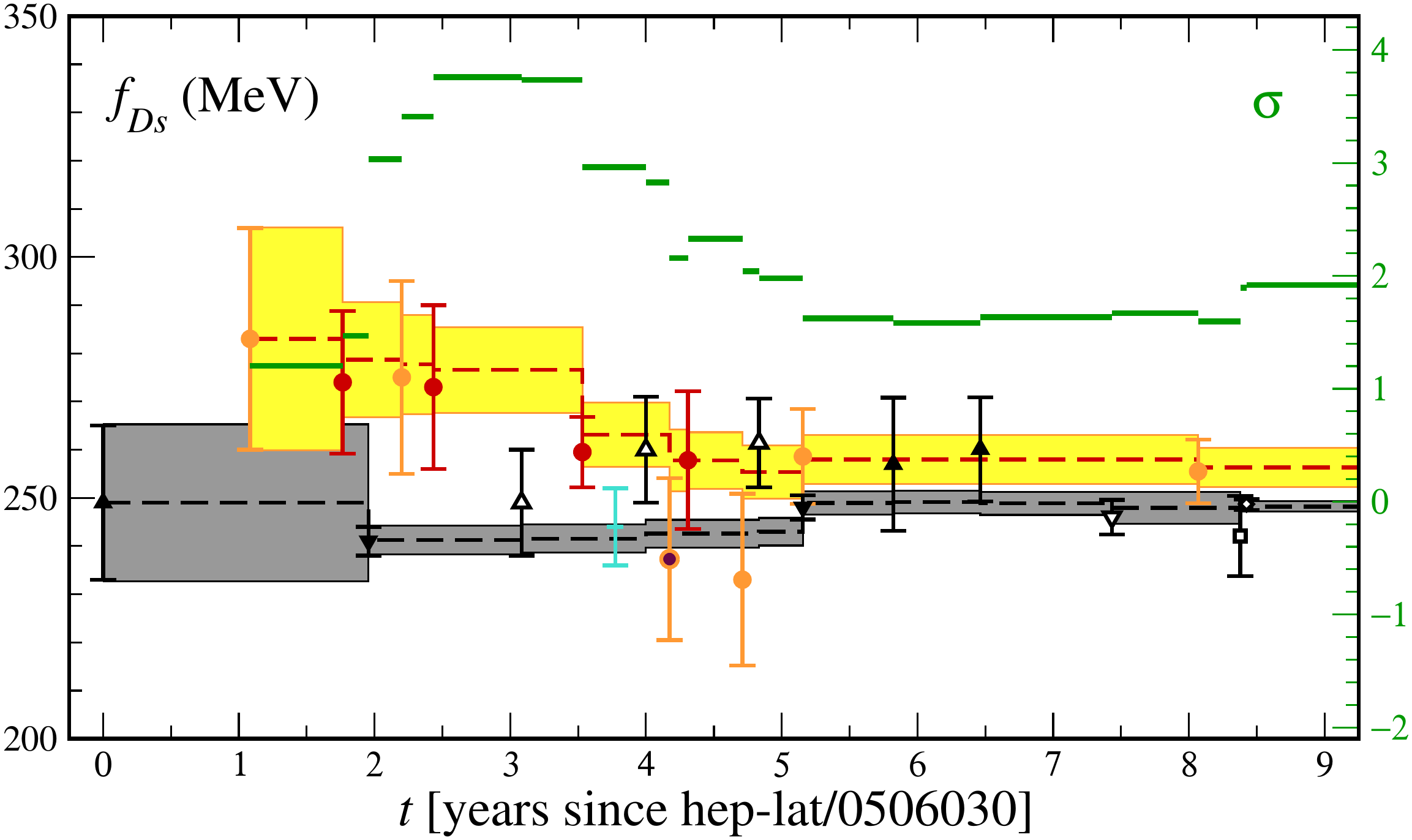}
    \caption{$f_{D_s}$ puzzle over the years.
        Author's averages of $f_{D_s}$ from lattice QCD (gray band) and $|V_{cs}|f_{D_s}$ from experiment, 
        divided by $|V_{cs}|$ from CKM unitarity (yellow band).
        Measurements from charm threshold (red); and $D_s$ in flight at the $\Upsilon(4S)$ (orange).
        For the calculations,  the number of sides of the symbols corresponds to $n_f$, the number of sea 
        quarks.
        Closed symbols are published; open are conference proceedings.
        Symbol rientation denotes charm formulation: Fermilab method denoted by up-pointing triangles;  
        staggered (HISQ) by down-pointing triangles ($n_f=2+1$) and diamonds ($n_f=2+1$);
        twisted-mass Wilson fermions by flat bars ($n_f=2$, cyan, not in the average) and squares ($n_f=2+1$).  
        The right axis and green curve denote the discrepancy in $\sigma$.
        See Refs.~\cite{Kronfeld:2009cf,Kronfeld:2012uk} for a full set of references.}
    \label{fig:fDs}
\end{figure}

Leptonic decays were supposed to yield a similar narrative, but did not.
Figure~\ref{fig:fDs} shows the history of unquenched lattice-QCD calculations, together with recent 
measurements~\cite{Kronfeld:2009cf,Kronfeld:2012uk}.
(Earlier measurements had very large error bars.) %
The first two with $\sim10\%$ errors \cite{Aubert:2006sd,Pedlar:2007za} agreed well with the first $n_f=2+1$
calculation (i.e., up, down, and strange quarks in the sea) \cite{Aubin:2005ar}.
In 2007, however, the first calculation of $f_{D_s}$ treating charm with light-quark
methods~\cite{Follana:2007uv} found a result that was significantly---nearly $4\sigma$ at one stage---lower
than experiment.
This discrepancy caused some consternation, including studies of how new physics could enter this process and
not others~\cite{Dobrescu:2008er,Dorsner:2009cu}; see also Ref.~\cite{Fajfer:2013bya}.

The history has been complicated by two notable shifts from normalization corrections.
In one case, HFAG~\cite{Asner:2010qj} proposed that the denominator of the relative branching ratio measured
in Ref.~\cite{Aubert:2006sd} could be improved, using a more suitable measure of $\phi\to KK$.
As a consequence the first orange point (near $t=1$) in Fig.~\ref{fig:fDs} moved down, to the one with a
maroon filling (near $t=4$).
Had this normalization been in place from the outset, no puzzle would have arisen.
In another case, the conversion from lattice units to MeV changed with the analysis of a wider set of data
(especially finer lattices), causing the step in the gray band near $t=5$.
The discrepancy, if it can be called that, now stands under $2\sigma$ (with $\sigma$ principally from
experimental statistics).

The total uncertainty for $f_D$ and $f_{D_s}$ from lattice QCD has now reached the level of
$0.5\%$~\cite{Na:2012uh,Dimopoulos:2013qfa,Bazavov:2013nfa}.
To push further---in kaon physics as well---we have to worry more about isospin and QED effects.
Isospin is principally a computing problem.
But how should one to connect a matrix element computed in finite volume to the physical photon cloud? %
Other effects are worth mentioning.
Electroweak box diagrams lead to a correction factor $1+(\alpha/\pi)\log(M_Z/\mu)$ for all leptonic and
semileptonic decays of hadrons~\cite{Sirlin:1981ie}.
In between the electroweak scale of and that of bremsstrahlung~\cite{Barberio:1993qi} lie electromagnetic
corrections that depend on hadron structure.
The interfaces between these scales should be handled with effective field theory, as it is for
kaons~\cite{Cirigliano:2008wn}, with lattice QCD+QED, or with a judicious combination.

\section{\boldmath$D^0$-$\littlebar{D}^0$ Mixing and Nonleptonic Decays}
\label{sec:nonleptonic}

At a superficial glance, $D^0$-$\littlebar{D}^0$ mixing is like mixing in other neutral meson systems,
$K^0$-$\littlebar{K}^0$, $B^0$-$\littlebar{B}^0$, and $B^0_s$-$\littlebar{B}^0_s$.
Indeed, short-distance contributions to $\Delta M_D$ and $\Delta\Gamma_D$---whether standard or BSM---follow
the same pattern.
In particular, any nonstandard model can be described by five matrix elements of $\Delta C=2$ operators.
Knowing the matrix elements can, thus, constrain extensions of the standard model~\cite{Golowich:2007ka}.
Lattice-QCD calculations of all five matrix elements are underway \cite{Chang:2013gla}.

Because, however, the neutral $D$ system mixes via diagrams with down-type quarks in the loop, long distance
effects, in which two $\Delta C=1$ interactions transpire at distances of order $\Lambda_\mathrm{QCD}^{-1}$
apart, can be just as large as the short distance contributions.
These effects are notoriously difficult to control: $m_c$ is too small for heavy-quark methods, while $M_D$
is too large for hadronic methods~\cite{Petrov:2003un}.
Note that lattice QCD has recently tamed the long-distance contribution to $\Delta M_K$~\cite{Christ:2012se}
with methods like those needed to tackle nonleptonic decays.

So far, lattice QCD has not played a significant role in (strong or weak) nonleptonic decays.
Nonexperts should read the next sentence carefully: %
The problem is not the lattice, i.e., the discrete spacetime; instead it arises from a conflict between
Euclidean space and finite volume~\cite{Maiani:1990ca}, two other choices necessary to place the problem on a
computer.
It would be wonderful to surmount this obstacle, because some of the most exciting physics is nonleptonic---%
for example the saga of $\mathcal{A}_\mathrm{CP}(D\to\pi\pi)-\mathcal{A}_\mathrm{CP}(D\to KK)$~\cite{Lenz:2013pwa}.
Until recently, progress in nonleptonic $D$ decays seemed hopeless (to me anyway), but now a recent spurt of 
activity provides some hope.

The foundation for the new work was developed over 25 years ago with L\"uscher's formalism for (nearly)
elastic problems such as $\pi\pi\to\pi\pi$ scattering \cite{Luscher:1986pf} and $\rho\to\pi\pi$
decay~\cite{Luscher:1990ux}.
Two notable extensions of this work have been to moving frames~\cite{Rummukainen:1995vs} and weak decays such
as $K\to\pi\pi$~\cite{Lellouch:2000pv}.
These have only recently become tractable in numerical lattice QCD~\cite{Dudek:2012xn}.

The two key insights are that the energy spectrum of a quantum field theory in a finite volume is discrete
and, moreover, that the energy levels are intimately related to the $S$~matrix.
Because the Euclidean-space evolution operator $\exp(-\hat{H}x_4)$ is just as suitable as $\exp(i\hat{H}t)$
for computing eigenvalues of the Hamiltonian $\hat{H}$, these insights allow us to use the favorite tool of
lattice gauge theory---the exponential fall-off of correlations functions---to access information about the
$S$ matrix.

Considerable mathematical physics (scattering theory on a torus) leads to a master formula.
If one partial wave dominates, the formula simplifies, and one directly obtains the $2\times2$ scattering
amplitude $\mathcal{M}_{2\to2}$ from
\begin{equation}
    F(E_n,\bm{P},L)\, \mathcal{M}_{2\to2}(\mathcal{E}_n) = -1,
    \label{eq:FM1}
\end{equation}
where $\mathcal{E}_n^2=E_n^2-\bm{P}^2$, and $E_n$ is the $n$th energy level of momentum $\bm{P}$ in a 
periodic box of size $L$.
Mathematics, not dynamics, provides the function $F$.
The algorithm, then, is to choose box size $L$ at the outset, pick several two-body 
$\bm{P}=2\pi(\bm{n}_1+\bm{n}_2)/L$, and compute the levels $E_n$ for these $\bm{P}$.
If a few partial waves matter, then one must resolve a few$\times$few determinant, but the basic structure 
of choosing $\bm{P}$, computing $E_n$, and plugging into a formula like Eq.~(\ref{eq:FM1}) still holds.

As mentioned above, the old formalism tackled only elastic (or nearly elastic) kinematics.
For nonleptonic $D$ decays, or for the long-distance part of $D^0$-$\littlebar{D}^0$ mixing, rescattering
effects play a role, however.
In the past year or so, many authors have been generalizing these methods, taking steps to understand
hadronic systems with more than two hadrons~\cite{Hansen:2012tf,Briceno:2012yi,Doring:2012eu}.
In particular, it is now known (conceptually) how to compute the 3$\times$3 scattering
amplitude~\cite{Polejaeva:2012ut,Kreuzer:2012sr,Briceno:2012rv,Hansen:2013dla}.
Weak amplitudes, needed for $D$ decay, can then be obtained from formalism for strong interactions via
perturbation in weak Hamiltonian~\cite{Lellouch:2000pv,Hansen:2012tf}.
All this ideas, and more (probably including effective field theories, as in Ref.~\cite{Beane:2007qr}) will
be needed before tackling processes such as $\pi\pi\to\pi\pi\pi\pi$ and $\pi\pi\to\pi\pi\pi\pi\pi\pi$, which
are a prerequisite to nonleptonic $D$ decays and long-distance $D$ mixing.
It is too early to see the light at the end of the tunnel, but at least the tunnel has been breached.

\section{Outlook}
\label{sec:outlook}

The LHC experiments have observed a particle that looks like the standard Higgs boson, but not the bevy of 
other particles anticipated by TeV-scale model builders.
Without such new states, the way forward in particle physics is through precision physics, such as rare 
and sensitive processes.
Precise experiments require commensurately precise theoretical calculations.
In my view, the past decade has seen lattice QCD move into this arena, down to the 1--2\% level.
Experiments studying charmed quarks played an important role in this enterprise.
That said, it seems to me that the time has come to use lattice QCD and experiment together, to understand 
physics better.

Indeed, lattice QCD has entered an era where the challenges lie in aspects that will be harder for 
experiments and other theoretical approaches to verify.
In some cases, small effects such as QED and isospin violation must be incorporated.
In other areas, precision is not yet paramount.
Rigorous treatment of many problems in charm physics will require a rigorous treatment of multi-hadron 
states.
This includes excited-state and exotic spectroscopy~\cite{Padmanath:2013bla,Prelovsek:2013cta},
as well as nonleptonic $D$ decays.


\Acknowledgements
I would like to thank Marco Gersabeck and Chris Parkes for their invitation, skillful organization, and kind 
hospitality during Charm~2013. 
I would also like to thank Jonna Koponen for sending information from Ref.~\cite{Koponen:2013tua} needed to 
prepare Fig.~\ref{fig:ffDK},
and Maxwell Hansen for conversations about nonleptonic $D$ decays.

\end{document}